\documentclass[letter]{ptptex}
\usepackage{wrapft}

\usepackage{color}

\usepackage{graphicx}

\pubinfo{, Submitted October 2, 2011}

\title{
Effects of a New Triple-$\alpha$ Reaction on the $S$-process in Massive Stars}

\author{
Yukihiro \textsc{Kikuchi}$^{1,}$\footnote{E-mail: kikuchi@phys.kyushu-u.ac.jp},
Masaomi \textsc{Ono}$^{1}$,
Yasuhide \textsc{Matsuo}$^{1}$
Masa-aki \textsc{Hashimoto}$^{1,}$\footnote{
E-mail: hashimoto@phys.kyushu-u.ac.jp},
and Shin-ichiro \textsc{Fujimoto}$^{2}$
}

\inst{
$^1$Department of Physics, Kyushu University, Fukuoka 812-8581, Japan
\\
$^2$Kumamoto National College of Technology, Kumamoto 861-1102, Japan
}

\abst{%
Effects of a new triple-$\alpha$ reaction rate on the $s$-process during the evolution of a
massive star of $25~M_{\odot}$ are investigated for the first time. Although the $s$-process in
massive stars has been believed to be established with only minor change,
we find that  the $s$-process with use of the new rate during the core He-burning is very 
inefficient compared to the case with the previous triple-$\alpha$ rate. However, the difference of
the overproduction is found to be largely compensated by the subsequent C-burning. Since the $s$-process
in massive stars has been attributed so far to the neutron irradiation during core He-burning,
our finding reveals  the importance of C-burning for the  $s$-process during
the evolution of massive stars.}

\PTPindex{240, 471}   

\begin{document}
\maketitle

\noindent
1. {\it Introduction\qquad}

A drastic change has been presented for the triple-$\alpha$ (3$\alpha$) reaction rate, which has been calculated by Ogata et al. (hereafter referred to OKK) \cite{rf:okk}
and found to become suddenly very large for the temperature below $2\times10^8$~K compared with the previous rate used so far.\cite{rf:nh88,rf:hashi95,rf:angulo}, and a rate by
 Fynbo et al.~\cite{rf:fynbo} (we call the reference as Fynbo) which revised the 3$\alpha$ rate of Ref.~\citen{rf:angulo} based on new experiments at high temperature of $T > 10^{9}$~K, where in the present study 
we can regard the differences between the two rates as unimportant.

As a consequence, the new rate results in the helium 
($^4$He) ignition in the lower density/temperature
on the stellar evolution of low-, intermediate-, and high-mass stars\cite{rf:dotter,rf:morel}, accreting white dwarfs\cite{rf:nom82,rf:nom82b}, and accreting neutron stars.\cite{fuji81,miya85,peng2010,matsuo2011}
Therefore, it is urgent to clarify quantitatively as possible as how the new rate affects the above astrophysical phenomena, because the rate plays the most fundamental role among the nuclear
burnings for almost all stars and could do some role in the early universe, where
any direct experiments for the 3$\alpha$ reaction are very difficult.  Related to the observations,
one of the most important impacts is concerned with the origin of the elements.~\cite{bbfh57}
Considering the evolution of massive stars, the $s$-process beyond iron (Fe) up to the mass number
$A\simeq 90$ has been studied
many times and confirmed as one of the site of nucleosynthesis;
The $s$-process is understood to be the neutron capture process for the production of heavier
nuclei up to Bi.\cite{bbfh57,kappler11} Usually the process is classified to two categories by the 
neutron densities $n_{\rm n}$ available to the $s$-process. One is
the weak component (${ 60 < A < 90}$) and the other is  the main component
(${ 90 < A < 205}$), where so called, pure 41 $s$-nuclei in the solar system abundances are
 known.~\cite{ag89}

 Prantzos, Hashimoto and Nomoto\cite{phn90} have confirmed that the $s$-process 
 occurs efficiently during core helium (He)-burning in massive stars in producing
the weak component of the $s$-elements. 
 Neutrons have been produced efficiently during core He-burning through the reaction ${\rm ^{22}Ne(\alpha,n)^{25}Mg}$. However, this reaction competes with the ${\rm ^{22}Ne(\alpha,\gamma)^{26}Mg}$ reaction. 
 Furthermore neutrons are quickly absorbed by the reactions of ${\rm ^{25}Mg(n,\gamma)^{26}Mg}$
and ${\rm ^{20}Ne(n,\gamma)^{21}Ne(n,\gamma)^{22}Ne}$. 
The neutron capture process after core He-burning has been studied by using a large network
as the post process calculations with use of the stellar models of $15 M_{\odot}$ and $30 M_{\odot}$
stars.\cite{ala91}  They have concluded that
no significant enhancement has occurred during core carbon (C)-burning compared to core
He-burning.  
Recently, effects of uncertainties of nuclear reactions on the $s$-process in massive stars 
have been examined in detail,
where we can find some changes in the overproduction accompanying uncertainties of the $3\alpha$ reaction
combined with $\rm^{12}C(\alpha,\gamma)^{16}O$ reaction~\cite{tur09}, and $\rm^{12}C+^{12}C$ reaction~\cite{hir10}.
Although these reaction rates are not determined definitely, the scenario of the $s$-process in massive
stars survives without objection at the present stage. On the other hand,
to follow the $s$-process
correctly, the appropriate nuclear reaction network must be used.
 In particular,  $s$-process branchings appear during the nucleosynthesis in
$T = 2 - 4 \times 10^8$ K for nuclei of $^{63}$Ni, $^{85}$Kr and $^{79}$Se.~\cite{kappler11}

In the present paper, we investigate the effects of  a newly calculated 3$\alpha$ reaction rate (OKK rate)\cite{rf:okk} on the production of the weak
component of the $s$-process (weak $s$-process)  in the evolution of a massive star of $25 M_{\odot}$ whose
helium core corresponds to $8 M_{\odot}$.


     \begin{figure}
     \centerline{\includegraphics[width=13 cm,height=8 cm] {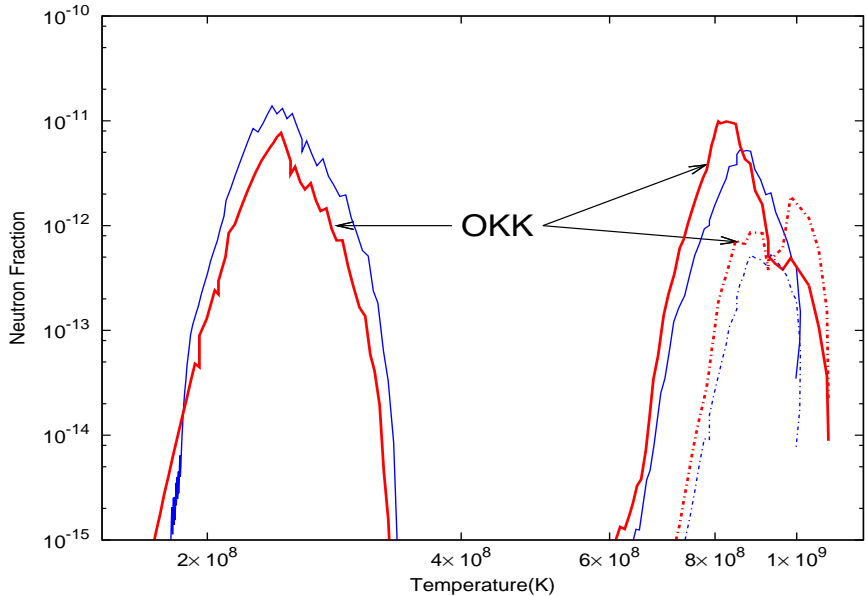}}                           
     \caption{Increased abundances of neutrons $\Delta y_{\rm n}$ against the temperature during core He-burning
 and core C-burning obtained at the center. 
Thin curves are the results with the use of the reaction rate of Fynbo.
Thick ones are the results  of OKK. Neutrons liberated by the reactions of 
$\rm^{22}Ne(\alpha,n)^{25}Mg$ are illustrated by the solid lines and those by
$\rm^{12}C+^{12}C$ are by the dotted lines. }
     \label{fig:1}
     \end{figure}

     \begin{figure}
     \centerline{\includegraphics[width=12 cm,height=7 cm] {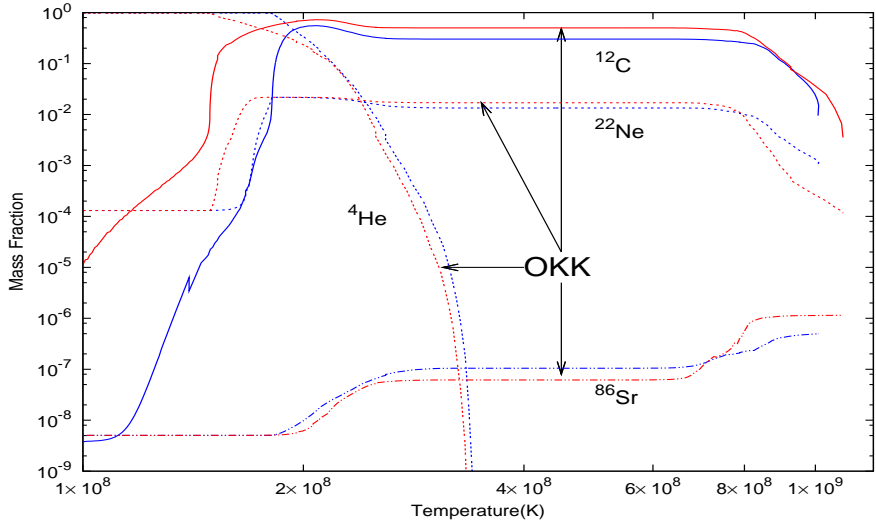}}
     \caption{Evolution of mass fractions of nuclei at the center responsible for the neutron production
against the temperature from core He-burning to the end of core C-burning. A typical pure-$s$ nucleus
is also shown.}
     \label{fig:2}
     \end{figure}

     \begin{figure}
     \centerline{\includegraphics[width=12 cm,height=8 cm] {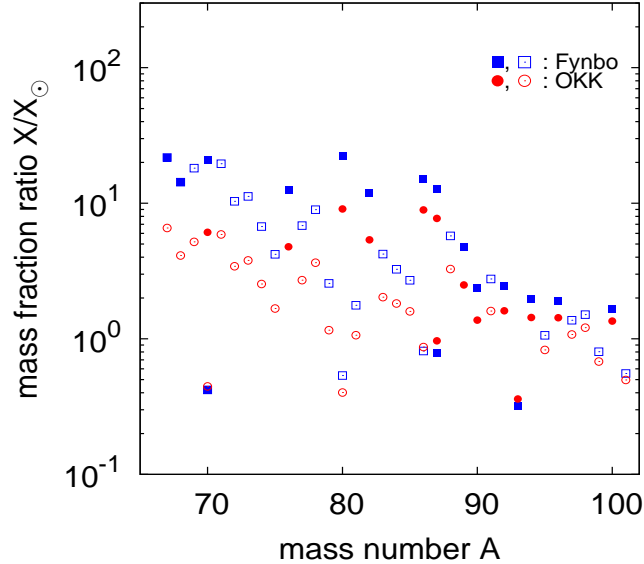}}                           
     \caption{Overproduction factors against the mass number $A$ at the end of core He-burning.
 Filled and open squares are the results with the use of reaction rate of Fynbo.
Dots and circles are the results  of OKK. Filled symbols correspond to the pure $s$-nuclei.}
     \label{fig:3}
     \end{figure}

     \begin{figure}
     \centerline{\includegraphics[width=12 cm,height=8cm] {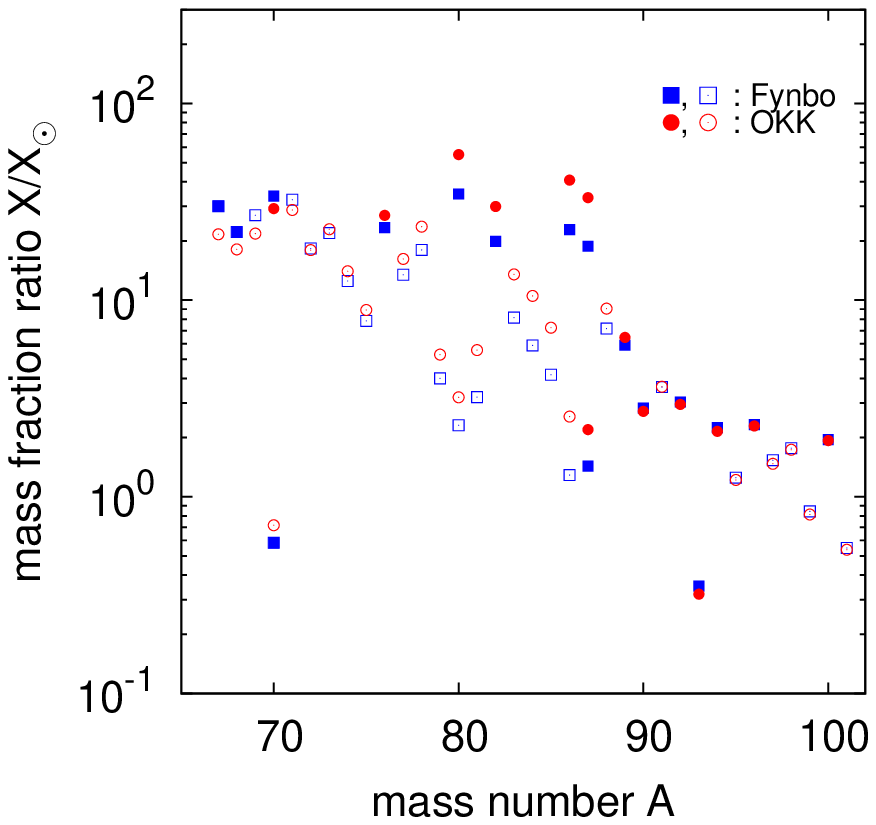}}                           
     \caption{Same as Fig.~\ref{fig:3} but for the epoch of the end of core C-burning. }
     \label{fig:4}
     \end{figure}

\noindent
2. {\it The $s$-process during  nuclear burnings\qquad}

 Since the weak $s$-process has been examined in detail for the star of $25~M_{\odot}$ in the main
 sequence mass of $M_{\rm ms}$~\cite{tur09}
and the mass around $25~M_{\odot}$ is regarded as an important star from the point of the
chemical evolution of galaxies,~\cite{kappler11}  we give the results of the $s$-process for the star.

We have developed  the nuclear reaction network based on the method by
 Prantzos, Arnould and Arcoragi\cite{paa87} of which network contains 440 nuclei from $^{12}$C to $^{210}$Bi.
 Our network extends to 1714 nuclei from neutron and proton to $^{214}$U linked with particle reactions and weak  interactions,\cite{nishi06,ono09} because some unknown reaction channels be open.
The reaction rates are taken from 
JINA REACLIB compilation,\cite{cyburt} where  new nuclear data has been included for the charged particle reactions and the ${\rm (n,\gamma)}$ cross sections after those of Bao et al.\cite{bao00} 
The $\beta$-decay and electron capture rates at finite temperature and density  are taken from  
Takahashi and Yokoi for nuclei above $^{59}$Fe.~\cite{ty87}  

The initial compositions are almost set to be the solar system abundances:~\cite{ag89} the original CNO elements
are converted to $^{14}$N, whose mass faction $X(^{14}\rm N)$ is 0.0137, and other elements heavier than oxygen are distributed in proportion to the solar values. The stellar evolutionary calculations are performed for the helium core of 
$M_{\alpha} = 8 M_{\odot}$ until the end of core C-burning as has been done in Refs.~\citen{rf:nh88,rf:hashi95}.

The extent of the weak $s$-process during the He-burning is determined by the size of the convection.
After the contraction of the helium core, convection begins from the center at the beginning of
 He-burning and ends when helium exhausts at the center.
 Since helium exhausts rather rapidly, convection disappears suddenly in response to the helium abundance.  
The size of the convection reaches the maximum of 5.28 $M_{\odot}$ for the case of Fynbo and
5.47 $M_{\odot}$ for OKK. 
The neutron source reaction ${\rm ^{22}Ne(\alpha,n)^{25}Mg}$ occurs efficiently near the end of core He-burning and the $s$-process proceeds to build up the nuclei beyond Fe. The neutron capture continues until
 core He-burning ends, where nuclei up to the mass number of $A \sim 90$ are overproduced notably.
This nucleosynthesis of the weak $s$-process has been proved many times by the $s$-process network calculations
combined with the stellar evolution calculations.\cite{paa87,phn90} 

After the end of He-burning at the center, gravitational contraction proceeds. Just after the ignition of
carbon at the center, convection begins from the center 
and ends at the epoch of carbon exhaustion.
After the size of the convection reaches the maximum of 0.79 (1.21 for OKK) $M_{\odot}$, 
it disappears as similar to the case of He-burning. 

In Fig.~\ref{fig:1} increased neutron abundances $\Delta y_{\rm n}$ for each epoch at the center are plotted as a function of the temperature, where the abundance $y_{\rm n} = n_{\rm n}/(\rho\cdot N_A)$ with the 
Avogadro number $N_A$ and the mass density $\rho$.
Since the abundance change occurs in different epochs for the two rates of Fynbo and
OKK, we choose the horizontal axis to be the temperature instead of time.
While neutrons during core He-burning
are liberated much more for the case of Fynbo compared to that of OKK,
the neutron abundances for the case of OKK have been much liberated during the active C-burning stage.  
These neutrons are supplied by the reaction of  ${\rm ^{22}Ne(\alpha,n)^{25}Mg}$, where  $\alpha$
particles are produced by $\rm^{12}C+^{12}C\longrightarrow$ $\rm^{20}Ne + \alpha$.
We note that the neutrons are supplied also to some extent  by $\rm^{12}C+^{12}C\longrightarrow$ $\rm^{23}Mg + n$
as shown by the dotted lines in Fig.~\ref{fig:1}.
 Clearly we can see the difference of the efficiency of the neutron productions for the two reaction rates. 
In Fig.~\ref{fig:2}, mass fractions responsible for neutron productions, $^{4}$He, $^{12}$C, $^{22}$Ne, and
a representative  $s$-element of $^{86}$Sr are plotted, where values of abundances correspond to those
of the central region. Let us compare their productions for OKK and Fynbo.
For the case of OKK,  $^{4}$He exhausts earlier than for Fynbo. Therefore, ${\rm ^{22}Ne(\alpha,n)^{25}Mg}$
is much more effective for Fynbo as seen in Fig.~\ref{fig:1} (first peaks for  $T\sim3\times10^8$~K ).
As the result, abundance of $^{12}$C for OKK compared to that for Fynbo becomes larger 
by a factor of 1.66 before C-burning begins. At the same time, it is found that $^{22}$Ne remains a lot, which produces neutrons significantly during core C-burning as seen in Fig.~\ref{fig:1} 
 (second peaks for $T\sim8\times10^8$~K). 
On the other hand, we can see the direct neutron productions by $\rm^{12}C+^{12}C\longrightarrow$ $\rm^{23}Mg + n$
for $T\sim(8-10)\times10^8$~K (third peaks). 
Effects of the production of $s$-elements can be recognized from the evolution of $^{86}$Sr, which is a pure
$s$-element. At the end of core He-burning, it is overproduced for Fynbo relative to that for OKK
by a factor of 2 until core
C-burning begins. However, the production is reversed in response to the second peak of the
neutron densities in Fig.~\ref{fig:1}. The production of $^{86}$Sr for OKK compared to that for Fynbo
 is increased by more than a factor of 2.3.

In Figs.~\ref{fig:3} and \ref{fig:4},
final mass fractions with respect to the initial ones, $X_i/X_{\odot}$, are plotted as a function of the mass number for all the stable nuclei. We note that the final mass fraction of the $i$-th element $X_i$ is obtained from 
the all layers of $8M_{\odot}$.
For the case of Fynbo,
large overproduction factors in the the range of  60 $< A < $ 90 are obtained at the end of core He-burning.  Above $A=90$, no significant enhancement can be seen as previous studies.~\cite{paa87,phn90}
On the contrary, the overproduction has been decreased  significantly for OKK. 
As a consequence, we find that the efficiency of the weak $s$-process during core He-burning stage
is very low if we use the OKK rate as seen in Fig.~\ref{fig:3} for all nuclei of $60 < A < 100$.
Since He-burning occurs much efficiently for the OKK rate, the amount of available helium to be captured by $^{22}$Ne becomes lower compared to the case of Fynbo.
Therefore, the results by OKK is inconsistent with the usual scenario of the weak
 $s$-process.~\cite{paa87,phn90}
 
 However, after core C-burning, overproduction for OKK surmounts that for Fynbo as seen in
Fig.~\ref{fig:4} and as discussed related to Fig.~\ref{fig:1} and/or Fig.~\ref{fig:2}. 
As the result, scenario concerning
the weak $s$-process which has been believed to occur mainly in He-burning stage
is changed if the OKK rate is adopted.

3. {\it Discussion and Conclusions\qquad}

     \begin{figure}
     \centerline{\includegraphics[width=12 cm,height=8cm] {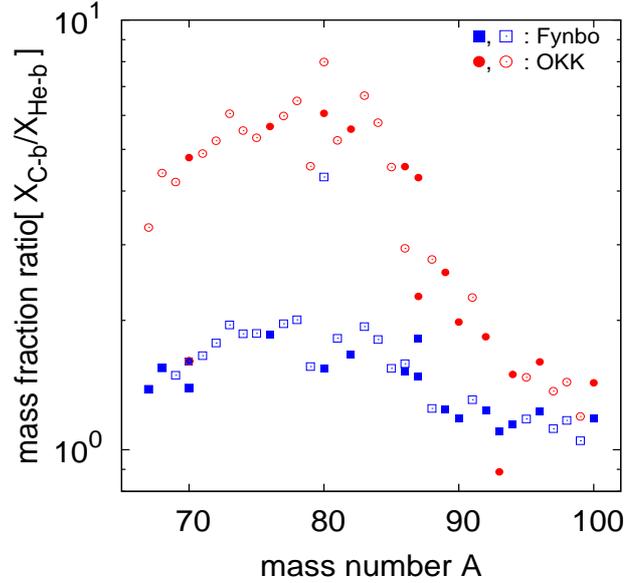}}                           
     \caption{Ratios of mass fractions of elements produced at the end of  two burning stages:   
 Mass fractions at the end of core C-burning divided by those at the end of
 core He-burning stages. The meaning of symbols are the same as in Fig.~\ref{fig:3}. }
     \label{fig:5}
     \end{figure}

We have found a possible new scenario of the weak $s$-process as follows.
During the He-burning stage, the weak component of the $s$-process
are produced  significantly for the Fynbo rate compared to those for the OKK rate (see Fig.~\ref{fig:3}).
 During the core
C-burning stage, neutron captures are activated much more for the OKK rate. As the result,
overproductions become larger than the case of Fynbo (see Fig.~\ref{fig:4}). 
To elucidate the degree of difference of the produced abundances at the end of core He- and
 C-burnings,
we show in Fig.~\ref{fig:5} the ratios of mass fractions calculated with the Fynbo and OKK rates, respectively;
the ratios are obtained from the mass fractions in Fig.~\ref{fig:4} divided by those in Fig.~\ref{fig:3}. 
We can recognize that the ratios are larger by a factor of 2-5 for the OKK rate compared to the case
of the Fynbo rate.
 This conclusion 
must have been maintained until the core collapse begins. Furthermore, the produced $s$-elements
would be preserved in the outer region of the explosively burning layer 
in the supernova ejecta.~\cite{kappler11}
 
The importance of carbon
burning stage on the weak $s$-process has been pointed out already comparing the uncertainties of the rate
of  $\rm^{12}C+^{12}C$.~\cite{hir10}  However, changes in the overproduction factors 
from helium to C-burning stage are not shown and/or discussed.  

Although we have studied the weak $s$-process of a $25~M_{\odot}$, qualitative effects induced from
the difference of the two reaction rates should remain for other massive stars.
While  stars of  $M_{\rm ms} < 20 M_{\odot}$ burn only a small fraction of $^{22}$Ne, enhanced
carbon would activate the $s$-process again.  Therefore, the weak $s$-process with the use of the OKK rate 
for relatively low massive stars
could contribute the chemical evolution of galaxies compared to the weak 
$s$-process considered  previously.~\cite{kappler11}

 While the weak component of the $s$-process are produced abundantly in either rates
for the solar values of the initial abundances, the
weak $s$-process  still remains uncertain due to the objectionable 3$\alpha$ reaction rate
coupled with other uncertain reaction rates.\cite{kappler11}
In case of low 
metallicity stars, the $s$-process during He-burning must be affected seriously by the products ${\rm ^{12}C, ^{16}O, ^{20}Ne}$, and $^{24}$Mg. 
How efficiently these 
nuclei become poisons depend on a relation between [O/H] and [Fe/O] as 
simulated in Ref.~\citen{phn90}, where for example $\rm [O/Fe]=\log_{10}[(X(O)/X(Fe)/(X(O)/X(Fe))_{\odot}]$.
Oxygen originates from the explosion of massive stars and would become a strong poison due to the enhanced
cross section.~\cite{nitmmu91}  On the other hand, if we adopt the OKK rate, carbon dominates oxygen
for massive stars of $M_{\rm ms}\le 25 M_{\odot}$. As a consequence, abundances of poisons will change 
star by star and therefore production of the $s$-components would be changed significantly compared to the
previous study.~\cite{phn90,rayet00}  In particular, in the low metallicity stars, the $p$-process elements which originate from the $s$-elements~\cite{rayet95}  could be produced more than previously
  by using the OKK rate. It is worthwhile to investigate the effects of the new 3$\alpha$ rate on the 
$p$-process in massive stars of $M\leq25M_{\odot}$.

\section*{Acknowledgements}
This work has been supported in part  by a Grant-in-Aid for Scientific Research 
(19104006, 21540272, 22540297) of the Ministry of Education, Culture, Sports, Science and 
Technology of Japan.


\begin{thebibliography}{99}
\bibitem{rf:okk} K. Ogata, M. Kan and M. Kamimura, \PTP{122,2009,1055}.\\
		K. Kan et al., JHP-Supplement-20 (1996), 204. \\ 
		K. Kan, Master thesis 1995 in Kyushu University, unpublished.
\bibitem{rf:nh88} K. Nomoto and M. Hashimoto, Phys. Rep. \textbf{163} (1988), 13.
\bibitem{rf:hashi95} M. Hashimoto, Prog. Theor. Phys. \textbf{94} (1995), 663. 
\bibitem{rf:angulo} C. Angulo et al., \JL{Nucl. Phys. A,656,1999,3}.
\bibitem{rf:fynbo} H. O. U. Fynbo et al., \JL{Nature,433,2005,136}. 
\bibitem{rf:dotter} A. Dotter and B. Paxton, \JL{Astron. Astrophys.,507,2009,1617}.
\bibitem{rf:morel} P. Morel, J. Provost, B. Pichon, Y. Lebreton and F. Th${\rm {\acute e}}$venin, \JL{Astron. Astrophys.,520,2010,A41}.
\bibitem{rf:nom82} K. Nomoto, \AJ{253,1982,798}.
\bibitem{rf:nom82b}  K. Nomoto, \AJ{257,1982,780}.
\bibitem{fuji81} M. Y. Fujimoto, T. Hanawa and S. Miyaji, \AJ{247,1981,267}.
\bibitem{miya85} S. Miyaji and K. Nomoto, \JL{Astron. Astrophys.,152,1985,33}.
\bibitem{peng2010} F. Peng and C. D. Ott, \AJ{725,2010,309}.
\bibitem{matsuo2011}  Y. Matsuo et al., ptp submitted (astro-ph/1105.5484).  
\bibitem{bbfh57}{E. M. Burbidge, G. R. Burbidge, W. A. Fowler and F. Hoyle,  Rev. of Modern Phys. {\bf 29} (1957), 547.}
\bibitem{kappler11} F. K\"{a}ppler et al. Rev. of Modern Phys., {\bf 83} (2011), 157.
\bibitem{ag89}{E. Anders and N. Grevesse, Geochim. Cosmochim. Acta {\bf 53} (1989), 197.}
\bibitem{phn90}{N. Prantzos, M. Hashimoto, K. Nomoto, Astron. Astrophy. {\bf 234    } (1990), 211.}
\bibitem{ala91}{J.-P. Arcoragi, N. Langer and M. Arnould, Astron. Astrophys. {\bf 249} (1991), 134.}
\bibitem{tur09} A. Heger and S. M. Austin, \AJ{702,2009,1068}.
\bibitem{hir10} M. E. Hirschi et al., Journal of Physics: Conference series, {\bf 202}, 2010, 012023.
\bibitem{paa87}{N. Prantzos, M. Arnould, and J.-P. Arcoragi, Ap. J. {\bf 315} (1987), 209.}
\bibitem{nishi06} S. Nishimura, et al., \AJ{642,2006,410}.
\bibitem{ono09} M. Ono, M. Hashimoto, S. Fujimoto, K. Kotake and S. Yamada, \PTP{122,2009,755}.
\bibitem{cyburt} Cyburt et al., \AJ{189,2010,240}
\bibitem{bao00}{Z. Y. Bao et al., Atomic Data and Nucl. Data Tables {\bf 76} (2000) 70.}
\bibitem{ty87}{K. Takahashi and K. Yokoi, Atomic Data and Nucl. Data Tables {\bf 36} (1987), 375.}
\bibitem{nitmmu91} T. Nagai, M. Igashira, K. Takeda, N. Mukai, S. Motoyama, F. Uesawa and H. Kitazawa, \AJ{372,1991,683}.
\bibitem{rayet00}  M. Rayet and M. Hashimoto, \JL{Astron. Astrophys.,354,2000,740}.
\bibitem{rayet95}  M. Rayet et al., \JL{Astron. Astrophys.,298,1995,517}.


\end{thebibliography}
\end{document}